\documentclass[11pt]{article}
\usepackage{euscript}
\usepackage{amssymb}
\usepackage{amsfonts}
\usepackage{amsbsy}
\usepackage{epsfig}
\usepackage{amsthm}
\usepackage{amscd}
\usepackage{amstext}
\usepackage{verbatim}
\usepackage{amsmath}
\usepackage{cancel}
\usepackage{capt-of}
\usepackage{empheq}

\usepackage[pdftex]{hyperref}

\def\ben{\begin{equation}}
\def\een{\end{equation}}

\let\pa=\partial
\def\be{\begin{equation}}
\def\ee{\end{equation}}
\def\beq{\begin{equation}}
\def\eeq{\end{equation}}
\def\ba{\begin{array}}
\def\ea{\end{array}}

\def\dalemb#1#2{{\vbox{\hrule height .#2pt
       \hbox{\vrule width.#2pt height#1pt \kern#1pt
               \vrule width.#2pt}
       \hrule height.#2pt}}}

\newcommand{\bea}{\begin{eqnarray}}
\newcommand{\eea}{\end{eqnarray}}
\newcommand{\tr}{{\rm tr} }

\def\Z{{{\Bbb Z}}}

\begin{document}

\begin{center}

{ \Large {\bf
Holographic mutual information and \\
distinguishability of Wilson loop and defect operators.
}}

\vspace{1cm}

Sean A. Hartnoll and Raghu Mahajan

\vspace{1cm}

{\small
{\it Department of Physics, Stanford University, \\
Stanford, CA 94305-4060, USA }}

\vspace{1.6cm}

\end{center}

\begin{abstract}

The mutual information of disconnected regions in large $N$ gauge theories with holographic gravity duals can undergo
phase transitions. These occur when connected and disconnected bulk Ryu-Takayanagi surfaces exchange dominance. That is, the bulk `soap bubble' snaps as the boundary regions are drawn apart. We give a gauge-theoretic characterization of this transition: States with and without a certain defect operator insertion -- the defect separates the entangled spatial regions -- are shown to be perfectly distinguishable if and only if the Ryu-Takayanagi surface is connected. Meanwhile, states with and without a certain Wilson loop insertion -- the Wilson loop nontrivially threads the spatial regions -- are perfectly distinguishable if and only if the Ryu-Takayanagi surface is disconnected. The quantum relative entropy of two perfectly distinguishable states is infinite.
The results are obtained by relating the soap bubble transition to Hawking-Page (deconfinement) transitions in the R\'enyi entropies, where defect operators and Wilson loops are known to act as order parameters.
\end{abstract}

\pagebreak
\setcounter{page}{1}

\section{Introduction and objective}

Holographic duality \cite{Maldacena:1997re} implies that the gravitational Hawking-Page transition is a deconfinement phase transition in a dual large $N$ gauge theory \cite{Witten:1998zw}. Deconfinement is the gauge theory dual of the gravitational dynamics {\it par excellence} --  the formation of black holes.

In recent years it has been understood that the gravitational thermodynamics of black hole event horizons can be substantially generalized by considering  entanglement entropy in the dual field theory \cite{Ryu:2006bv, Lewkowycz:2013nqa}. The entanglement entropy of a spatial region in the `boundary' field theory is universally given by the area of a bulk minimal surface that ends on this boundary region. Black hole entropy arises as a special case. These minimal surfaces provide a direct probe of the bulk geometry.

Consider the entanglement entropy of two disconnected regions in the quantum field theory. In theories with holographic gravity duals, there are two different candidates for the minimal surface whose area will give the entanglement entropy. Firstly there are disconnected bulk surfaces that end on the two boundary regions separately. Secondly there may be a single bulk surface that connects the two boundary regions through the bulk. The two possibilities are illustrated in figure \ref{fig:con} below. When the boundary regions are close together, the connected bulk surface typically has lower area and `wins', whereas when the boundary regions are far apart the disconnected bulk surfaces dominate. This is a rather robust phase transition in the bulk `soap bubble'. It manifests itself in the fact that the mutual information between the two regions vanishes (to leading order at large $N$ \cite{Barrella:2013wja, Faulkner:2013ana}) in the case where the disconnected surfaces dominate.
\begin{figure}[h]
\centering
\includegraphics[height = 0.13\textheight]{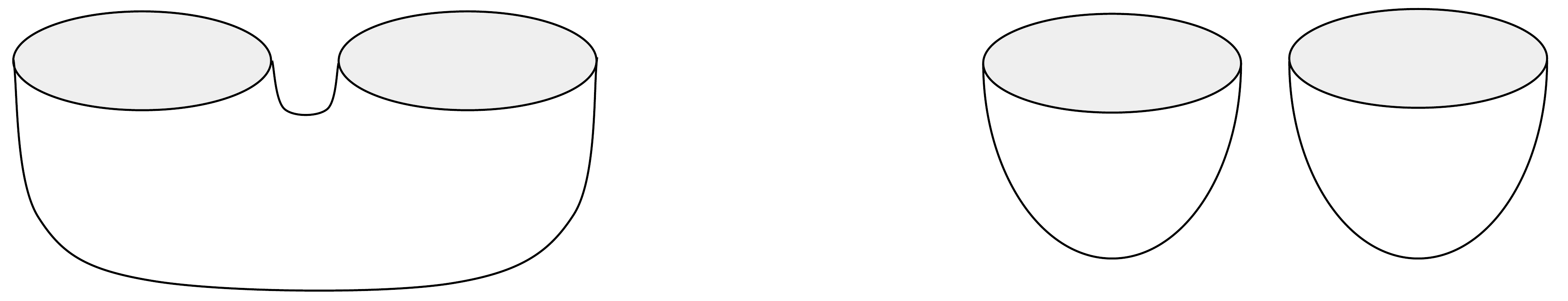}
\caption{\label{fig:con} {\bf Connected and disconnected bulk minimal surfaces}, ending on a disconnected boundary region.}
\end{figure}

Our objective is to give a simple gauge-theoretic characterization of the phase transition in the bulk minimal surfaces. The main results are equations (\ref{eq:final}), (\ref{eq:main2}) and (\ref{eq:final2}) below. Equation (\ref{eq:main2})  is the more fundamental of these statements. The statements involve the reduced density matrices formed with and without certain defect operator and Wilson loop insertions. The defect operator separates the two components of the entangled spatial regions, while the Wilson loop operator nontrivially links (i.e. threads) them. We show that the reduced density matrices with and without the defect operator insertion are perfectly distinguishable if and only if the Ryu-Takayanagi surface is connected. Meanwhile, the reduced density matrices with and without the Wilson loop insertion are perfectly distinguishable if and only if the Ryu-Takayanagi surface is disconnected. Perfectly distinguishable density matrices have infinite relative entropy. Thus in theories with classical gravity duals, the vanishing or not of the mutual information can be captured in a precise gauge-theoretic way. Our strategy will be to relate the `soap bubble' phase transition to the Hawking-Page transition, following \cite{Headrick:2010zt}. 

\section{R\'enyi entropies and Hawking-Page transitions}
\label{sec:renyi}

The reduced density matrix associated to a spatial region $A$ of a quantum field theory in, say, the vacuum is given by the
Euclidean functional integral
\be\label{eq:rho}
\rho[\phi_+(x),\phi_-(x)] = \frac{1}{Z} \int {\mathcal D}\phi \, e^{-S[\phi]} \delta\Big(\phi_A(0^+,x)-\phi_+(x)\Big)
\delta\Big(\phi_A(0^-,x)-\phi_-(x)\Big) \,.
\ee
Here $\phi_A(t,x)$ is the restriction of the field $\phi$ to the region $A$ at time $t$. The reduced density matrix is a function of the field data on the region $A$.

The R\'enyi entropies are
\be\label{eq:renyi}
S_n \equiv \frac{1}{1-n} \log \tr \rho^n \,,
\ee
and the entanglement entropy
\be
S = \lim_{n \to 1^+} S_n = - \tr \left( \rho \log \rho \right) \,.
\ee
The R\'enyi entropies are equal to the partition function of the theory on an $n$-fold cover of the original space. In particular
\be\label{eq:rhon}
\tr \rho^n = \frac{Z_n}{Z_1^n} \,.
\ee
Here $Z_1 = Z$ is the partition function on the original space.

\begin{figure}[h]
\centering
\includegraphics[height = 0.23\textheight]{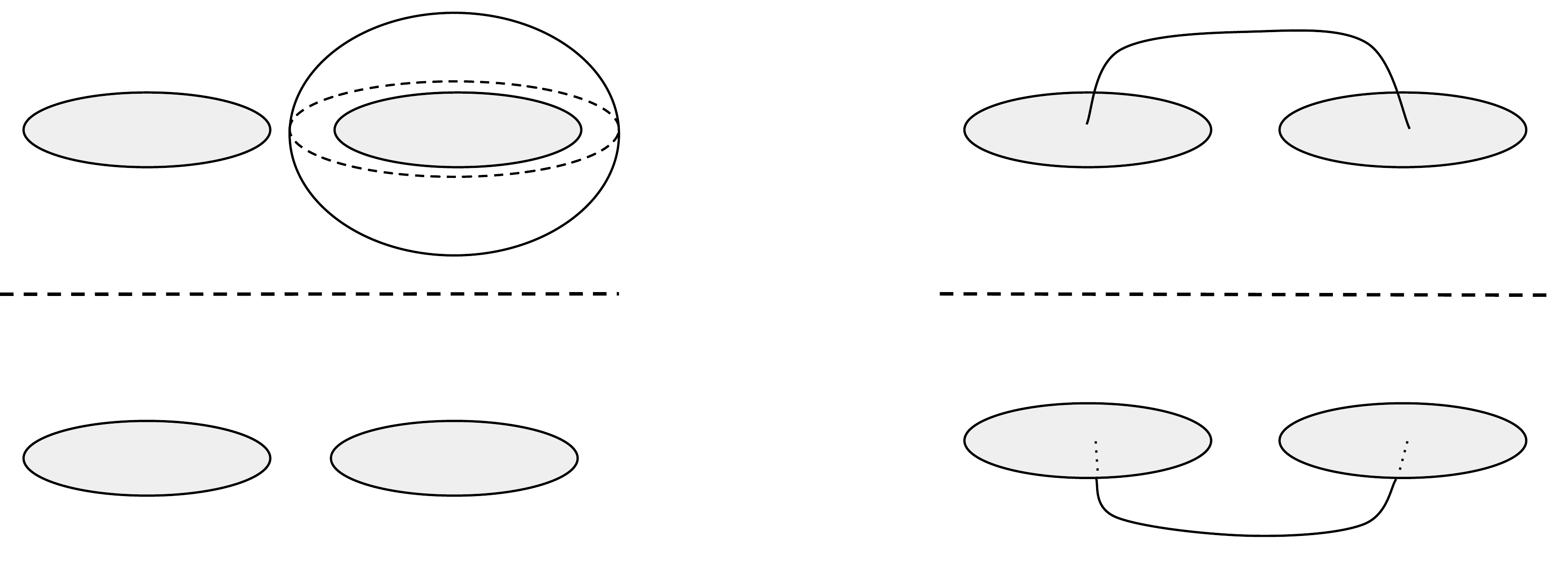}
\caption{\label{fig:cycles} {\bf Nontrivial cycles on the $n$-fold cover}. For illustration we have taken a two fold cover (above and below the central dashed line) of three dimensional spacetime with the entangled region given by two disconnected $d=2$ dimensional discs (shaded). The different sheets of the cover are glued together across these discs. The left hand figure illustrates the nontrivial 2-cycle. The right hand figure illustrates the nontrivial 1-cycle.}
\end{figure}

In the $n$-fold cover, $n$ copies of the original spacetime are glued together along the region $A$. Let us consider the case in which $A$ is given by two disconnected $d$-dimensional spatial regions in a $d+1$ dimensional spacetime. In this case, the $n$-fold cover is a topologically nontrivial spacetime. In particular, there are $n-1$ nontrivial $d$-cycles (we can think of these as $d$-spheres that enclose one of the regions on each sheet) and also $n-1$ nontrivial $1$-cycles (which we can think of as circles transversing the two regions). These cycles are illustrated in figure \ref{fig:cycles}. In counting the nontrivial cycles, each sheet is compactified at infinity.

\begin{figure}[h]
\centering
\includegraphics[height = 0.22\textheight]{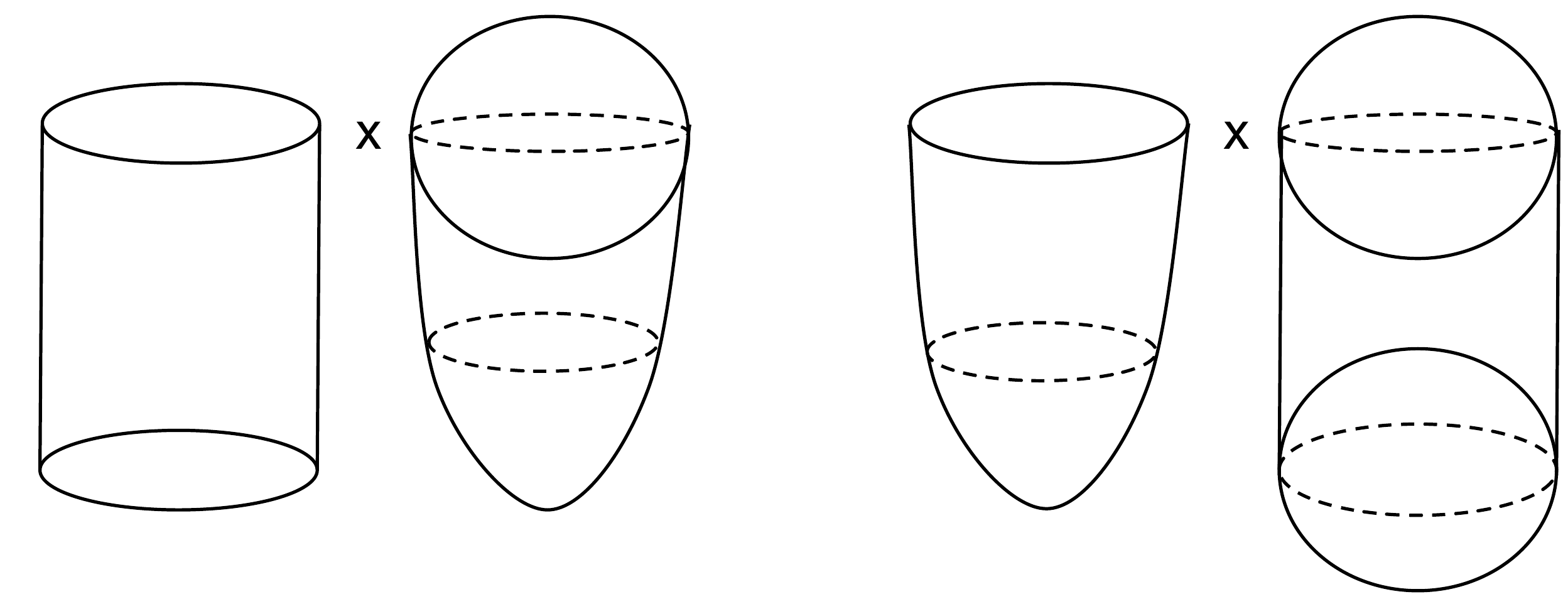}
\caption{\label{fig:HP} {\bf The Hawking-Page transition}. Two possible bulk topologies for an $S^1 \times S^2$ boundary geometry. In the left hand figure, the $S^1$ remains topologically nontrivial in the bulk whereas the $S^2$ is contractible (thermal AdS$_4$). In the right hand figure, the $S^1$ contracts but the $S^2$ is nontrivial (Schwarzschild-AdS$_4$).}
\end{figure}

In (\ref{eq:rhon}) the $n$th R\'enyi entropy is expressed as the partition function of the gauge theory on the $n$-fold cover, which we have seen is topologically nontrivial. To evaluate this partition function holographically in the semiclassical large $N$ limit, one needs to find the bulk solution whose boundary manifold is the $n$-fold cover geometry. When the boundary is topologically nontrivial, there are distinct topological possibilities for the bulk, depending on whether the various nontrivial boundary cycles are contractible in the bulk or not. This is illustrated in figure \ref{fig:HP} below for the case where the boundary is $S^1 \times S^2$, the original Hawking-Page transition \cite{Hawking:1982dh}. With one nontrivial 1-cycle and one nontrivial 2-cycle, this boundary geometry is topologically the same as the double cover 
geometry illustrated above in figure \ref{fig:cycles} (although the double cover geometry has singularities at the boundary of the region $A$).

When a nontrivial boundary 1-cycle is trivialized in the bulk, then one can consider a Euclidean fundamental string worldsheet wrapping the corresponding bulk `cigar' geometry (i.e. a surface in the bulk whose only boundary is the boundary 1-cycle). This string has a finite renormalized on shell action. A string worldsheet ending on the boundary computes the expectation value of a Wilson loop operator in the dual gauge theory \cite{Maldacena:1998im, Rey:1998ik}. When the 1-cycle is contractible in the bulk, the Wilson loop around the boundary 1-cycle is nonzero. If the 1-cycle remains nontrivial in the bulk, then no such finite action string worldsheet exists and the Wilson loop is zero. This Wilson loop is therefore an order parameter for the bulk topology \cite{Witten:1998zw}.

To apply an analogous argument to a nontrivial boundary $d$-cycle in the $n$-fold cover, we need a $d+1$ dimensional bulk object that can wrap a corresponding $d+1$ dimensional bulk `cigar' when the boundary cycle is contractible in the bulk. While less generically defined than the Wilson loop, the required objects typically exist in theories with holographic duals. Specifically, the $d+1$ dimensional boundary field theory describes the low energy excitations of a stack of coincident $Dd$-branes (possibly originating from higher dimensional branes wrapped on internal cycles). Euclidean $Dd$-branes can therefore be used in the bulk, and end on $d$-cycles on the boundary. In the boundary, this corresponds to inserting $d$-dimensional defect operators \cite{Karch:2000gx, DeWolfe:2001pq}. The defect operators are higher dimensional (for $d>1$) cousins of 't Hooft loops, e.g. \cite{Gukov:2006jk}. Following the logic of the previous paragraph, these operators are field-theoretic order parameters for the bulk topology. In particular, if the boundary $d$-cycle is not contractible in the bulk, then the expectation value of the defector operator on the cycle will vanish.

For the special case in which the boundary theory is $1+1$ dimensional, then we can use Wilson loops -- dual to fundamental strings -- as order parameters for both types of boundary cycle.

So far we have described two types of phase transitions: Hawking-Page type transitions, illustrated in figure \ref{fig:HP}, that can occur in the R\'enyi entropies, and phase transitions in the holographic entanglement entropy, illustrated in figure \ref{fig:con}. It was argued in \cite{Headrick:2010zt} that the `soap bubble' transitions in the entanglement entropy are in fact the $n \to 1^+$ limit of the Hawking-Page transitions. This connection can be shown for 1+1 dimensional CFTs with gravity duals by explicitly constructing the bulk geometries dual to the $n$-fold covers, in a way that allows continuation to non-integer $n$ \cite{Faulkner:2013yia, Hartman:2013mia}. It suggests that an order parameter for the entanglement entropy transitions can be obtained by taking an appropriate $n \to 1^+$ limit of the Wilson loop and D-brane order parameters for the R\'enyi transitions. This will be seen to be true. We will also see, however, that the fundamental underlying fact, distinguishability of certain density matrices, can be deduced from the integer R\'enyi entropies, without needing any analytic continuation.

\section{Distinguishability of defect operators}
\label{sec:relative}

Consider first the case of $d$-dimensional defect operators. The Wilson loop case will be described in the following section. We start by writing down the density matrix associated to the region A in the presence of such an operator, call it $W_\Sigma$, where $\Sigma$ is the worldvolume of the Euclidean defect:
\be\label{eq:rhoW}
\rho^{W_\Sigma}[\phi_+(x),\phi_-(x)] = \frac{1}{\langle W_\Sigma \rangle} \int {\mathcal D}\phi \, W_\Sigma \, e^{-S[\phi]} \delta\Big(\phi_A(0^+,x)-\phi_+(x)\Big)
\delta\Big(\phi_A(0^-,x)-\phi_-(x)\Big) \,.
\ee
The expectation value $\langle W_\Sigma \rangle_1 = \langle W_\Sigma \rangle$ appearing in the normalization is nonzero, there is no topology with a single sheet.
While the reduced density matrix  (\ref{eq:rho}) is obtained from the vacuum of the field theory,
the reduced density matrix described by (\ref{eq:rhoW}) corresponds to the state created by the (Lorentzian analytic continuation of the) defect operator insertion in the path integral. For instance, in the $1+1$ dimensional case in which $W_\Sigma$ can be taken to be a Wilson loop operator, it is the state created by dragging a heavy external quark along the (analytic continuation of the) worldline $\Sigma$. The worldvolume $\Sigma$ should be symmetric about the $t=0$ time slice.

An order parameter for a phase transition in the R\'enyi entropies is, as we discussed above, the expectation value of the defect operator around a nontrivial $d$-cycle in the $n$-fold cover. We take the 
defect worldvolume to surround one component of the region $A$ on only one of the $n$ sheets, for instance as illustrated in figure \ref{fig:cycles}. We will write this operator as $W_\Sigma^{(1)}$, by which we mean that the worldvolume $\Sigma$ is entirely on the first sheet.
This expectation value can conveniently be normalized as
\be\label{eq:us}
W_n \equiv \frac{Z_1}{\langle W_\Sigma \rangle_1} \frac{\langle W_\Sigma^{(1)} \rangle_n}{Z_n} =  \frac{\tr \left(\rho^{W_\Sigma} \rho^{n-1} \right)}{\tr \rho^n} \,.
\ee
For all integer $n > 1$, this quantity is zero when the boundary $d$-cycle does not contract in the bulk and nonzero if it does. We are interested in the analytic continuation to $n \to 1^+$, note that $W_1 = 1$ always, and therefore consider
\bea
- \lim_{n \to 1} \frac{d}{dn} W_n & = & \tr \left[ \left(\rho - \rho^{W_\Sigma} \right) \log \rho \right] \nonumber \\
& = & \Delta \langle H \rangle \;\;  = \;\;  S(\rho^{W_\Sigma} \parallel \rho) \; + \; \Delta S \,.\label{eq:orderone}
\eea
In the final line we expressed the quantity in two ways. Firstly, as the change in the expectation value of the modular Hamiltonian: $\rho \equiv e^{-H}/\tr(e^{-H})$. So $ \Delta \langle H \rangle = \tr ( \rho^{W_\Sigma} H ) - \tr ( \rho H)$. Secondly, we introduced the relative quantum entropy of $\rho^{W_\Sigma}$ with respect to $\rho$,
\be
S(\rho^{W_\Sigma} \parallel \rho) \equiv \tr \left(\rho^{W_\Sigma} \log \rho^{W_\Sigma} \right) - \tr \left(\rho^{W_\Sigma} \log \rho \right) \,,
\ee
as well as the difference in entanglement entropy upon insertion of the Wilson loop,
\be
\Delta S = S(\rho^{W_\Sigma}) - S(\rho) \, .
\ee
A useful discussion of quantum relative entropy can be found in \cite{Blanco:2013joa}. Klein's inequality states that $S(\rho^{W_\Sigma} \parallel \rho) \geq 0$, vanishing only when $\rho = \rho^{W_\Sigma}$. Both the relative entropy and the difference in entanglement entropy are UV finite quantities.

In the remainder we show that the quantity $S(\rho^{W_\Sigma} \parallel \rho) \; + \; \Delta S$ in (\ref{eq:orderone}) indeed captures holographic phase transitions in the mutual information -- of the type illustrated in figure \ref{fig:con}. We will show that the topology of the Ryu-Takayanagi minimal surface as $n \to 1^+$ is determined by the topology of the spacetimes that compute the R\'enyi entropies at integer $n$. This ties together the Hawking-Page transitions to the entanglement entropy transition. We must also explain how to analytically continue the holographic computation of the defect operator $\langle W_\Sigma^{(1)} \rangle_n$ to non-integer $n$.

Independently of how the defect operators are analytically continued in detail -- to be described below -- a simple argument shows that if two density matrices satisfy $\tr \left( \rho^A \rho^B \right) = 0$, then $\tr \left( \rho^A (\rho^{B})^{x} \right) = 0$ for all $x > 0$ and furthermore $S(\rho^A || \rho^B)$ is infinite. Work in a basis in which $\rho^B$ is diagonal, so that $\rho^B = \text{diag}(\lambda_1, \lambda_2, \ldots)$. Then $\tr \left( \rho^A \rho^B \right) = \sum_i \rho^A_{ii} \lambda_i = 0$. The positive semi-definite property of density matrices implies that $\lambda_i, \rho^A_{ii} \geq 0$ for all $i$. Thus for every $i$ either $\rho^A_{ii}$ or $\lambda_i$ vanishes. It follows that $\tr \left( \rho^A (\rho^B)^x \right) = \sum_i \rho^A_{ii} \lambda^x_i = 0$ for all $x>0$. Furthermore, the fact that $\tr \rho^A = 1$ requires at least one of the $\rho^A_{ii}$ to be nonzero. The corresponding $\lambda_i$ must therefore vanish. It follows that $\sum_i \rho^A_{ii} \log \lambda_i$, and hence $S(\rho^A || \rho^B)$, is infinite. Applying this logic to our case we have
\be\label{eq:infinite}
\langle W_\Sigma^{(1)} \rangle_n = 0 \quad \Rightarrow \quad \tr \left(\rho^{W_\Sigma} \rho^{n-1} \right) = 0 \quad \Rightarrow \quad S(\rho^{W_\Sigma} \parallel \rho) = \infty \,.
\ee
The change in entanglement entropy in (\ref{eq:orderone}) is finite, as we recall after equation (\ref{eq:final}) below. Therefore, it is the infinite relative entropy in (\ref{eq:infinite}) that carries the signature of the vanishing R\'enyi entropies $W_n$ with defect operator insertions to $n \to 1^+$. Very na\"ively we can think of this as the infinite negative derivative of $\frac{dW_n}{dn}$ at $n=1$ as it jumps from $W_1 = 1$ to $W_n = 0$ for $n > 1$. However, the conclusion (\ref{eq:infinite}) connects vanishing defect operators at integer $n$, e.g. $\tr \left(\rho^{W_\Sigma} \rho \right) = 0$ at $n=2$, directly to the relative entropy without need of analytic continuation.

A slightly stronger statement follows from $\tr \left(\rho^{W_\Sigma} \rho \right) = 0$. Because both density matrices are positive semi-definite Hermitian matrices, we can write $\rho^{W_\Sigma} = A A^\dagger$ and $\rho = B B^\dagger$, for some Hermitian matrices $A$ and $B$. It follows that $\tr \left[(A^\dagger B)^\dagger (A^\dagger B) \right] = 0$. This last statement implies $A^\dagger B = 0$ as a matrix equation. Consider a basis in which $\rho^{W_\Sigma}$ and hence $A$ are diagonal. The result $A^\dagger B = 0$ now implies that $B_{ik} = 0$ for all $i$ in the support of the diagonalized $A$. Hermiticity of $B$ then gives $B_{ki} = 0$. It follows that $\text{supp}\, A \subset \ker B$. Therefore $A$ and $B$, and hence $\rho^{W_\Sigma}$ and $\rho$, have orthogonal support.

Orthogonal support implies the infinite relative entropy of equation (\ref{eq:infinite}), but is a stronger condition in general. An important consequence of orthogonal support is that the density matrices are perfectly distinguishable. Consider a measurement described by POVM elements corresponding to (i) projection onto the support of $\rho^{W_\Sigma}$, (ii) projection onto the support of $\rho$, and (iii) projection onto the intersection of the kernels of $\rho^{W_\Sigma}$ and $\rho$. If the state is given by $\rho^{W_\Sigma}$ then the first element is observed with probability one whereas if the state is given by $\rho$, then the second element is observed with probability one. The third element is never observed. Therefore:
\be\label{eq:distinguishable}
\tr \left(\rho^{W_\Sigma} \rho \right) = 0 \quad \Rightarrow \quad \text{$\rho^{W_\Sigma}$ and $\rho$ perfectly distinguishable}.
\ee
The result (\ref{eq:distinguishable}) is well known (and elementary) in the context of quantum information theory. We have included a derivation here for completeness.

\section{Distinguishability of Wilson loops}
\label{sec:wilson}

In section \ref{sec:renyi} we noted that the $n$-fold cover geometries that compute the R\'enyi entropies for two disjoint spatial regions typically had nontrivial $1$-cycles and $d$-cycles. See figure \ref{fig:cycles}. The previous section \ref{sec:relative} has discussed the nontrivial $d$-cycles. We now return to the $1$-cycles.

The $1$-cycles can be wrapped by Wilson loop operators. Wilson loops are more universally defined than the defect operators we have discussed so far. On the other hand, as we see in figure \ref{fig:cycles}, the nontrivial $1$-cycles necessarily straddle more than one of the sheets. At first sight this might appear to complicate expressing the expectation value of the Wilson loop as a trace involving density matrices. For instance, consider a Wilson loop along a nontrivial curve $C$ that is supported on the first two of $n$ sheets (with $n \geq 2$), as illustrated in figure \ref{fig:cycles}. We can write this as
\be\label{eq:notsonice}
\frac{\langle W_C \rangle_n}{Z_n} = \frac{\langle W_+ \rangle_1 \langle W_- \rangle_1}{Z_1^2} \frac{\tr \left(\rho^+ \rho^- \rho^{n-2} \right)}{\tr \rho^n} \,.
\ee
Here $\rho^\pm$ are defined analogously to the $\rho^{W_\Sigma}$ that we considered previously in (\ref{eq:rhoW}). However, the difference is that now the Wilson loop $W_C$ must be split up into two parts $W_\pm$. The Wilson line $W_+$ is the part of $W_C$ on the first sheet while $W_-$ is the part of $W_C$ on the second sheet. This means that $\rho^\pm$ are not generally positive semidefinite density matrices because they are not constructed from a state times its hermitian conjugate. Instead, requiring that $W_+$ is obtained from $W_-$ by reflecting about the $t=0$ time slice,
we must write
\be\label{eq:DC}
\rho^+ = C D^\dagger \,, \qquad \rho^- = D C^\dagger \,,
\ee
for matrices $C$ and $D$. Therefore $\rho^\pm = \left( \rho^\mp \right)^\dagger$. This representation follows from writing the pure states with and without the $W_+$ insertion as
\be
| \text{vac} \rangle = \sum_{ij} D_{ij} |i\rangle_A |j\rangle_{A^\text{c}} \,, \qquad | W^+ \rangle = \sum_{ij} C_{ij} |i\rangle_A |j\rangle_{A^\text{c}} \,,
\ee
and then tracing over the compliment $A^\text{c}$ of the entangled region $A$ to obtain
\be
\rho^+ = \tr_{A^\text{c}} | W^+ \rangle \langle \text{vac} | \,, \qquad \rho^- = \tr_{A^\text{c}} | \text{vac} \rangle \langle W^+ | \,.
\ee

Unlike the defect operator case in (\ref{eq:us}), the expression (\ref{eq:notsonice}) is not naturally expanded about $n=1$. However, a more useful perspective exists. Define the density matrices
\be\label{eq:DD}
\hat \rho = \tr_A | \text{vac} \rangle \langle \text{vac} |= ( D^\dagger D )^T \,, \qquad \hat \rho^{W_C}  = \tr_A | W^+ \rangle \langle W^+ | = (C^\dagger C)^T \,. 
\ee
Note that these density matrices are defined by integrating over the degrees of freedom {\it inside} rather than outside the spatial region $A$. As we illustrate in the following figure \ref{fig:switch}, if we build the R\'enyi entropies associated to these density matrices, then the Wilson loop is now on purely the first sheet.
\begin{figure}[h]
\centering
\includegraphics[height = 0.22\textheight]{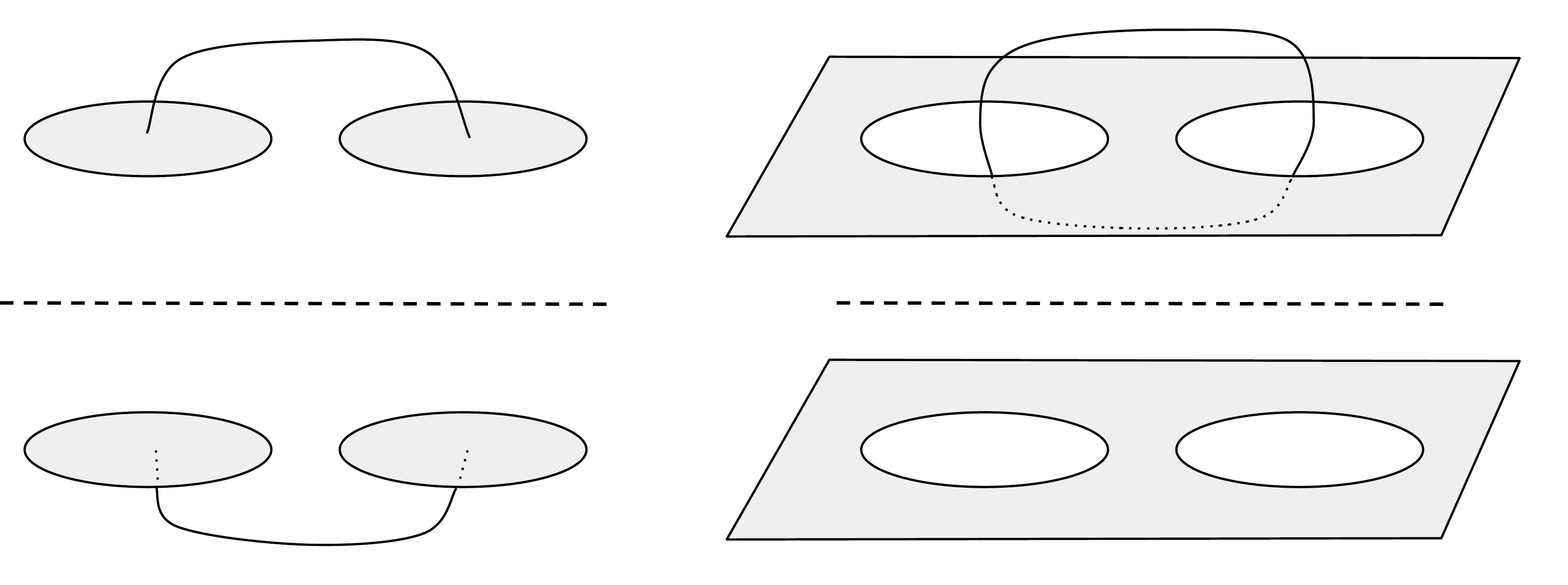}
\caption{\label{fig:switch} {\bf The double cover geometry for two different density matrices}. Left: The double cover geometry for the density matrix inside the region $A$ (the two disks). The Wilson loop spans the two sheets. Right: The double cover geometry for the density matrix outside the two disks. The Wilson loop is now entirely on the first sheet. In both cases the sheets are glued along the shaded grey regions.}
\end{figure}
In terms of these new density matrices, we can compute the expectation values of Wilson loops on the first sheet of an $n$-fold cover using $\langle \hat W_C \rangle_n \sim \tr  \left( \hat \rho^{W_C} \hat \rho^{n-1} \right)$. These formulae are now entirely analogous to those in (\ref{eq:us}) for defect operators, and the $n \to 1^+$ limit can be taken in exactly the same way.

Throughout this paper we will stick with the $n$-sheeted covers discussed in earlier sections, associated with the density matrices $\rho$ and $\rho^{W_\Sigma}$ and built by integrating out degrees of freedom outside the spatial region $A$. However, the previous paragraph shows that by instead integrating out degrees of freedom inside the spatial region $A$, exactly the same logic can be applied to the density matrices $\hat \rho$ and $\hat \rho^{W_C}$. Even using the `wrong' covers for the discussion of Wilson loops, we can obtain the desired result concerning distinguishability of $\hat \rho$ and $\hat \rho^{W_C}$.
Suppose that the boundary 1-cycle in the $n=2$ cover does not trivialize in the bulk. Then we have
\be
\langle W_C \rangle_2 = 0 \quad \Rightarrow \quad \tr \left(\rho^+ \rho^- \right) = 0 \quad \Rightarrow \quad \tr \left(\hat \rho \, \hat \rho^{W_C} \right) = 0\,.
\ee
The final step uses the definitions (\ref{eq:DC}) and (\ref{eq:DD}).
But $\hat \rho$ and $\hat \rho^{W_C}$ are now honest, positive semi-definite, density matrices. Therefore, by the arguments of the previous section we can conclude that when $\langle W_C \rangle_2$ vanishes, then $\hat \rho$ and $\hat \rho^{W_C}$ have orthogonal support, are perfectly distinguishable, and have infinite relative entropy.

Taking the present and previous sections together, we have shown that:
\bea
\text{$1$-cycles trivialized in bulk} \quad \Rightarrow \quad \rho, \rho^{W_\Sigma} \; \; \text{perfectly distinguishable} \,, \label{eq:a} \\
\text{$d$-cycles trivialized in bulk} \quad \Rightarrow \quad \hat \rho, \hat \rho^{W_C} \; \; \text{perfectly distinguishable} \,. \label{eq:b}
\eea
Once again, these results can be obtained directly from expectation values of Wilson loop and defect operators on the double cover geometry. The next step is to relate which cycle is trivialized to the connectedness of the Ryu-Takayanagi surface.

\section{Topology of the Ryu-Takayanagi surface}

The boundary geometry, prior to inserting the defect operator, admits a $\Z_n$ `replica' symmetry that rotates the $n$ planes of the cover. Following \cite{Lewkowycz:2013nqa} we assume that this symmetry extends to the bulk spacetime $M_n$. While unproven, this assumption has so far led to correct physical answers.\footnote{While the boundary geometry is  an $n$-fold cover of the $n=1$ geometry, this is typically not the case in the bulk. I.e. the bulk is not simply $n$ copies of the $n=1$ spacetime \cite{Headrick:2010zt, Lewkowycz:2013nqa}.} We can therefore take the quotient to obtain the spacetime
\be
\widetilde M_n \equiv M_n/\Z_n \,.
\ee
The spacetime $\widetilde M_n$ has a conical defect in the bulk along the fixed point locus of the $\Z_n$ action. On the asymptotic boundary of $M_n$, the fixed points of the $\Z_n$ action are the boundaries of the spatial region $A$ (the regions over which the $n$ sheets are glued). We can see this by considering a small loop that links the boundary of the spatial region $A$. This loop needs to encircle the boundary of the region $n$ times in order to close. We can parametrize this circle by a coordinate $\tau$ with period $2 \pi n$. The $\Z_n$ quotient then acts by identification $\tau \sim \tau + 2\pi$. These circles shrink to zero at the boundary of the spatial region $A$, which is therefore a fixed point locus of $\Z_n$. The $(d-1)$-dimensional fixed point locus of the boundary extends to a $d$-dimensional conical defect in $\widetilde M_n$. The key observation of \cite{Lewkowycz:2013nqa} is that the space $\widetilde M_n$ with a conical defect can be defined for any $n$, not just integers, and furthermore that as $n \to 1^+$, the conical defect surface becomes precisely the Ryu-Takayanagi minimal surface in the $n=1$ bulk spacetime. We proceed to argue that the topology of the conical defect surface in $\widetilde M_n$ -- and hence the Ryu-Takayanagi surface in the limit $n \to 1^+$ -- is determined by the topology of the bulk spacetime $M_n$. The result of this section is shown in figure \ref{fig:final} and equations (\ref{eq:s1}) and (\ref{eq:sd}) below.

To illustrate the general argument, we consider first in some detail the case of two disjoint intervals in 1+1 dimensions. In particular, for the case of a CFT, it is convenient (following e.g. \cite{Lewkowycz:2013nqa, Casini:2011kv}) to perform a Weyl transformation on the $n$-fold cover boundary geometry that sends the (four) fixed points to infinity. This renders the boundary geometry smooth and easier to visualize. Figure \ref{fig:abcd} below illustrates this map.
\begin{figure}[h]
\centering
\includegraphics[height = 0.35\textheight]{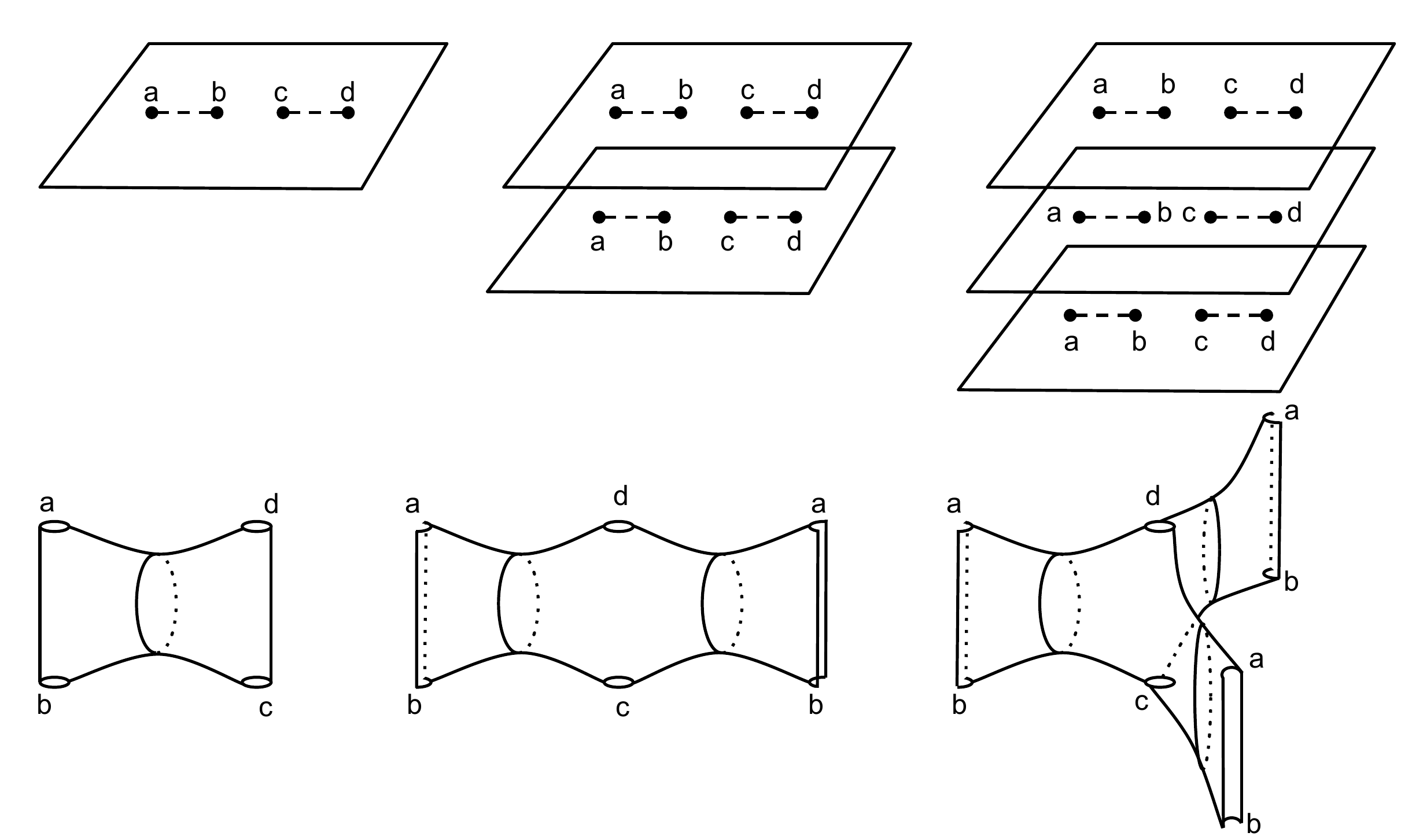}
\caption{\label{fig:abcd} {\bf Weyl transformation} of the $n$-fold cover (for $n=1,2,3$) of two intervals in 1+1 dimensions to a space in which the fixed points of the $\Z_n$ action (labelled $a,b,c,d$) are mapped to infinity. These spaces are respectively a sphere, torus and genus two Riemann surface, all with four points deleted. The boundaries joining $a$ and $b$ need to be sewn together in the obvious way, namely, the vertical lines are to be pairwise identified.}
\end{figure}

Consider in particular the $n=2$ geometry. The boundary geometry is $\Z_2$ symmetric. We can shade half of the boundary; this half is mapped into the other half by the $\Z_2$ action. See figure \ref{fig:z2} below.
The bulk is also $\Z_2$ symmetric (by assumption), and hence it must be possible to extend this shading into the bulk. The shaded and non-shaded regions of the boundary in figure \ref{fig:z2} are separated by two circles. One circle goes through the points $c$ and $d$ and the other goes through the points $a$ and $b$. These two circles must therefore extend into the bulk as boundaries of the shaded and non-shaded halves of the bulk. Thus there must be a surface $S$ in the bulk whose boundary $\partial S = S^1 \cup S^1$. This bounding surface is itself identified under the $\Z_2$ action, and will contain the line of fixed points/conical defects. By constraining the topology of this surface we will constrain the topology of the line of fixed points.
\begin{figure}[h]
\centering
\includegraphics[height = 0.1\textheight]{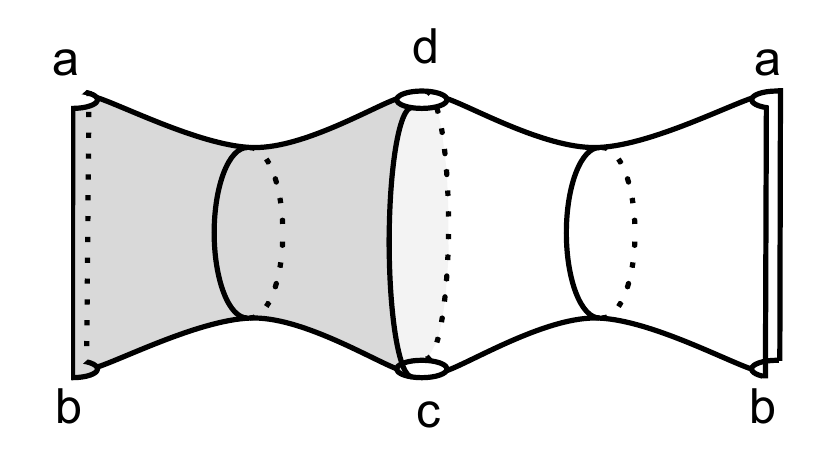}
\caption{\label{fig:z2} {\bf Division of the boundary geometry}, for $n=2$, into halves that are mapped into each other by the $\Z_2$ action. One half has been shaded. The boundary of the shaded region is two circles, one going through $a$ and $b$ and the other through $c$ and $d$. These circles are themselves each identified under the $\Z_2$ action, with $a,b,c,d$ as fixed points.}
\end{figure}

There are two topologically distinct ways to fill in the boundary torus, corresponding to trivializing different boundary cycles. We can think of this as either filling in the inside or the outside of the torus. The filling will be determined dynamically and transitions between the two possible fillings as a function of the modular parameter of the boundary torus are Hawking-Page transitions. Consider first the case in which the interior of the torus in figure \ref{fig:z2} is to be filled in. We must therefore determine the possible embeddings of the bounding surface $S$ (separating two halves of the interior of the torus that are to be mapped under the $\Z_2$ action). The simplest possibility is that $S = D_2 \cup D_2$, i.e. it is the union of two discs with boundaries given by the two circles described in the previous paragraph. This corresponds to the obvious shading of the bulk in which we shade half of the interior of the torus. The alternative possibility for the bounding surface is $S = S^1 \times [0,1]$. These two possibilities are illustrated in figure \ref{fig:z2fill} below.
\begin{figure}[h]
\centering
\includegraphics[height = 0.1\textheight]{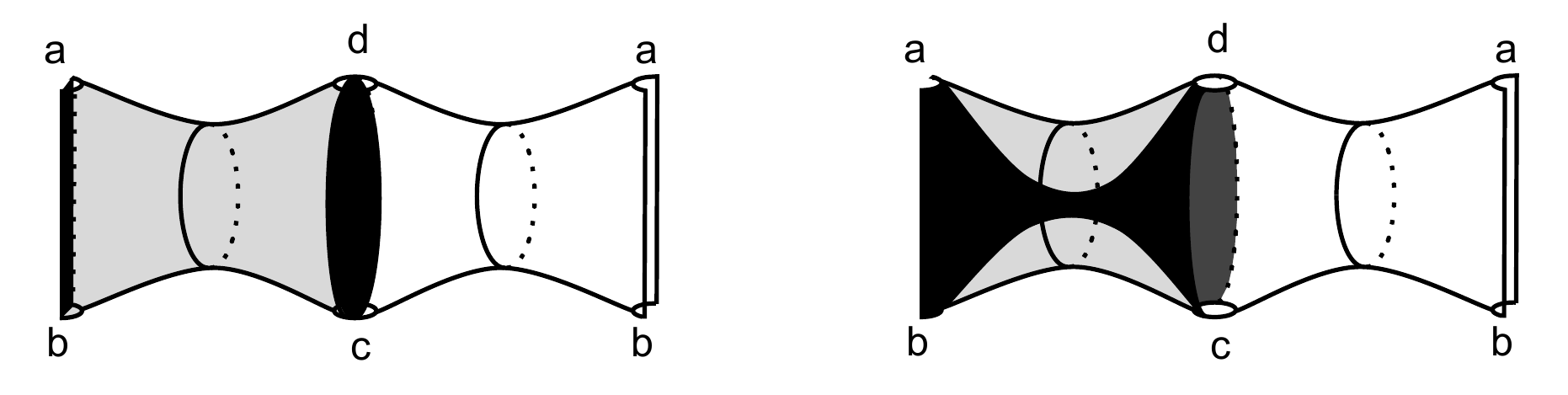}
\caption{\label{fig:z2fill} {\bf Two possible extensions of the boundary shading to the bulk}. The surface $S$ separating the bulk (a filled in torus) into two parts is shown in black. In the left figure $S = D_2 \cup D_2$. In the right figure $S = S^1 \times [0,1]$.}
\end{figure}

Of the two possibilities shown in figure \ref{fig:z2fill}, the surface $S = S^1 \times [0,1]$ cannot in fact separate the bulk into two halves that are mapped into each other under an extension of the boundary $\Z_2$ symmetry. Consider the two loops on the boundary that are shown in the figure (the two circles that separate $a$ and $d$ on either side of $d$). These are mapped into each other under the $\Z_2$ action on the boundary. However, under the division of the bulk by the surface $S = S^1 \times [0,1]$, one of these loops is contractible within its corresponding bulk region while the other is not (this argument is made before taking the $\Z_2$ quotient of the bulk). Therefore the homotopic contractions of the two loops cannot be mapped into each other, and hence this particular division of the bulk cannot correspond to an extension of the $\Z_2$ symmetry on the boundary. Only the other other division, with $S = D_2 \cup D_2$ is admissible.

We have just seen that there is a topologically unique extension of the $\Z_2$ symmetry into the bulk. The conical defect lines are contained within the bounding surface $S$. Under the $\Z_2$ action, each disc $D_2$ is folded on itself, leaving a line of fixed points running along the diameter of the disc. It follows that there are two lines of fixed points in the bulk: one running from the boundary fixed point $a$ to the boundary fixed point $b$ and another running from $c$ to $d$. Therefore we have fixed the topology of the bulk conical defect lines for $n=2$ and for a filled in torus.

The argument is immediately extended to the case of the $n$-fold cover. Here we shade the boundary in $n$ different colors. The extension of the $\Z_n$ symmetry, and hence the coloring, to the bulk is again topologically unique and given by the obvious generalization of the $n=2$ case. A slight difference in making the argument for the $n>2$ cases is that one must keep track of the line of fixed points itself in the bulk, as this is where multiple surfaces (each analogous to $S$ above) separating differently shaded regions must meet. The fixed point lines in the bulk must again run from $a$ to $b$ and from $c$ to $d$.

\begin{figure}[h]
\centering
\includegraphics[height = 0.1\textheight]{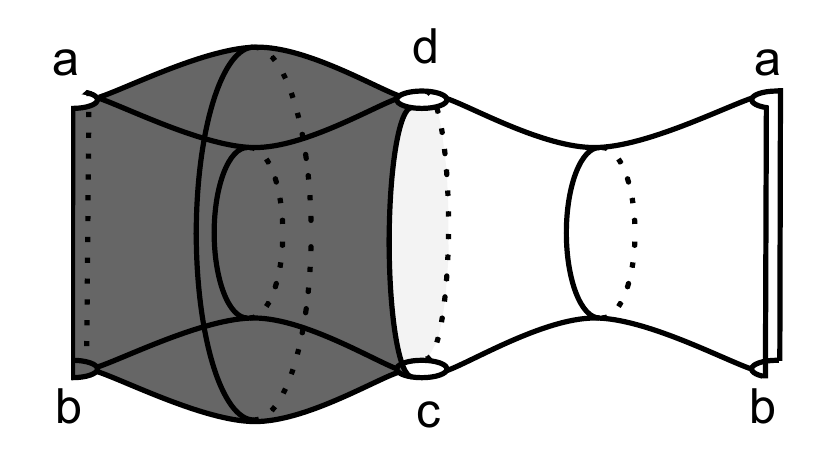}
\caption{\label{fig:z2fillout} {\bf Extension of the boundary shading to the bulk} in the case when the bulk is the exterior of the torus. The bounding surface is shown in dark grey and has topology $S = S^1 \times [0,1]$.}
\end{figure}

We can now consider the case in which the $n=2$ torus is `filled out' rather than `filled in' (i.e. the opposite boundary cycle is trivialized in the bulk). In this case the only possible bounding surface extending the boundary $\Z_2$ action into the bulk (i.e. the outside of the torus) has topology $S = S^1 \times [0,1]$. This surface is illustrated in figure \ref{fig:z2fillout}.
However, there is now no topological argument excluding this surface from separating two regions that are mapped into each other under a $\Z_2$ action. Indeed, a bulk surface with topology $D_2 \cup D_2$ does not exist in this case. Therefore, in this case, the fixed point lines in the bulk, which lie in the surface $S$, must connect $a$ to $d$ and $b$ to $c$.\footnote{One might also think of connecting $a$ to $c$ and $b$ to $d$ in this case. This, however, would lead to intersecting geodesics in the $n \to 1^+$ limit.
Arguments along the lines of those in \cite{Headrick:2007km} show that the intersecting geodesics will not be minimal. Therefore it is the configuration described in the main text that is relevant.} Note that the $\Z_2$ again acts on $S$ itself, so the only true boundary to $\widetilde M_2 = M_2/\Z_2$ will be the lines of fixed points. As before, this argument can be extended to $n>2$.

In the above arguments it is important that, whichever the filling, the boundary cycles have been trivialized in a $\Z_n$ invariant way. This allows us to avoid considering more complicated bulk geometries beyond the simple `interior' or `exterior' fillings (see e.g. \cite{Faulkner:2013yia}). This will be important in the following section when we compute the expectation value of certain defect operators. The two $\Z_n$ symmetric fillings are of course more invariantly characterized by which cycles on the $n$-fold cover are trivialized in the bulk (i.e. the $1$-cycles or the $d$-cycles).

The Ryu-Takayanagi minimal surface is the $n \to 1^+$ limit of the lines of conical defects \cite{Lewkowycz:2013nqa}. We must make the assumption that, for a given boundary geometry and choice of regions to entangle, the two possible bulk geometries (interior and exterior fillings) do not exchange dominance as a function of $n$. This is known to hold for the entanglement of two intervals in $1+1$ dimensions \cite{Faulkner:2013yia, Hartman:2013mia}. If there is indeed no change in filling as a function of integer $n$, the topology of the defect lines in $\widetilde M_n$ can be unambiguously continued to $n \to 1^+$. In particular we can conclude that the Ryu-Takayanagi surface will be connected when the exterior fillings dominate and disconnected when the interior fillings dominate. This is the main result of this section so far and is in agreement with explicit computations in \cite{Headrick:2010zt, Faulkner:2013yia}. We illustrate the relation in figure \ref{fig:final} below.
\begin{figure}[h]
\centering
\includegraphics[height = 0.125\textheight]{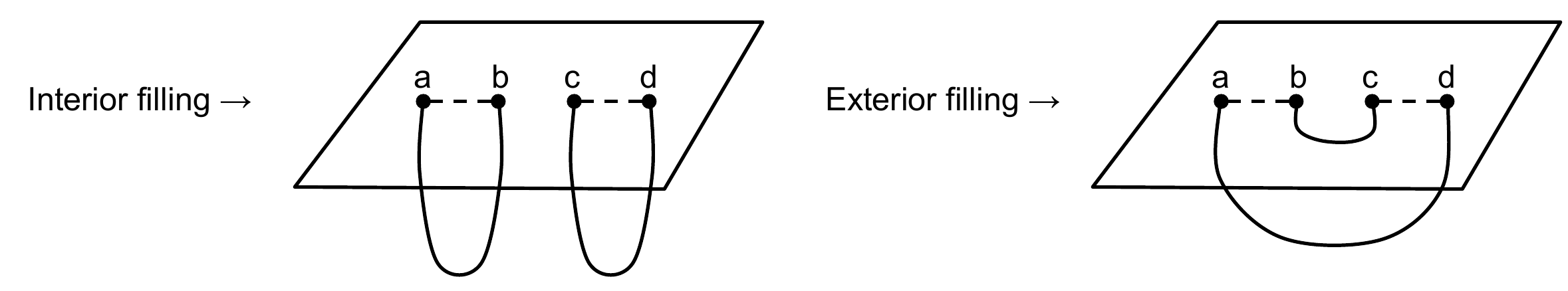}
\caption{\label{fig:final} {\bf The filling of the boundary $n$-fold covers} determines the topology of the Ryu-Takayangi surface.}
\end{figure}

Theories in general higher dimensions follow the same logic. We will restrict ourselves to the case in which there are two disconnected spatial regions, each with the topology of a disc $D_{d}$ in the $d+1$ boundary spacetime dimensions. As above, we can start by considering the $n=2$ geometry. The topology of the $n=2$ cover of the boundary geometry is $S^1 \times S^{d}$. This can be seen by considering an $S^1$ that transverses the spatial regions as shown on the right hand side of figure \ref{fig:cycles}. The geometry is then swept out by an $S^d$ at each point on the $S^1$. These $S^d$s become increasing pancake-like as they approach the disc regions. For the case of a CFT, it is again natural to perform a Weyl transformation that maps the conical singularities to infinity or otherwise `unflattens' the $S^d$s at the disks.

As in the $1+1$ dimensional case, discussed in detail above, the $\Z_2$ action divides the boundary $S^1 \times S^d$ geometry into two halves that are mapped into each other. These halves are divided by $S^d \cup S^d$, generalizing the $S^1 \cup S^1$ that we encountered before. Once again the question is how to extend this division into the bulk. Extending this division requires us to identify an appropriate bulk surface $S$ with boundary $\partial S = S^d \cup S^d$.

The geometry $S^1 \times S^d$ is the boundary of both $D_2 \times S^d$ and $S^1 \times D_{d+1}$. These are the two possible fillings. Exactly the same logic as previously leads to the conclusions that
\bea
\lefteqn{\text{$S^1$ filled in} \; \Rightarrow \; \text{bulk bounding surface $S = [0,1] \times S^d$}} \nonumber \\
&& \qquad \qquad \qquad \Rightarrow \; \text{conical defect surface connected}. \label{eq:s1} \\
\lefteqn{\text{$S^d$ filled in} \; \Rightarrow \; \text{bulk bounding surface $S = D_{d+1} \cup D_{d+1}$}} \nonumber \\
&& \qquad \qquad \qquad \Rightarrow \; \text{conical defect surface disconnected}. \label{eq:sd}
\eea

The discussion of the previous few paragraphs extends straightforwardly to the higher cover case with $n>2$. While the geometry is no longer globally a direct product, it is still locally (almost everywhere) of the form an interval times $S^d$. This is again seen by considering a one dimensional curve of $S^d$s that foliate the cover. For instance, for the case of $n=3$ the geometry can be described as a figure 8 curve of $S^d$s. For $d=1$ this is just the usual way one would draw a genus two Riemann surface. One possible filling is clearly to fill in the $S^d$s everywhere. In this case one recovers the conclusion in (\ref{eq:sd}), now generalized to all $n \geq 2$. The other filling is again the `exterior' filling that trivializes the fundamental group. This filling is seen to generalize (\ref{eq:s1}) to all $n$.

\section{Results}
\label{sec:results}

The remaining step is to give a holographic definition of the expectation value of defect operators and Wilson loops on nontrivial cycles that is suitable for analytic continuation to non-integer $n$. The analytically continuation is achieved following the ideas in \cite{Lewkowycz:2013nqa}. We describe the analytic continuation in the appendix, focussing on the defect operator case. This shows that the required analytic continuation exists. Having defined an analytic continuation of $\langle W_\Sigma^{(1)} \rangle_n/Z_n$ to non-integer $n$, we can take the $n \to 1^+$ limit described in equation (\ref{eq:orderone}) above.  However, we will not need the explicit continuation to establish our results. 

Consider first the defect operator case.  We know that the expectation value $\langle W_\Sigma^{(1)} \rangle_n$ of the defect operator will be zero if and only if the $d$-cycle that it wraps is not contractible in the bulk. This occurs when the boundary $n$-fold cover is filled `out' rather than in, as illustrated in figure \ref{fig:defect} below.
\begin{figure}[h]
\centering
\includegraphics[height = 0.1\textheight]{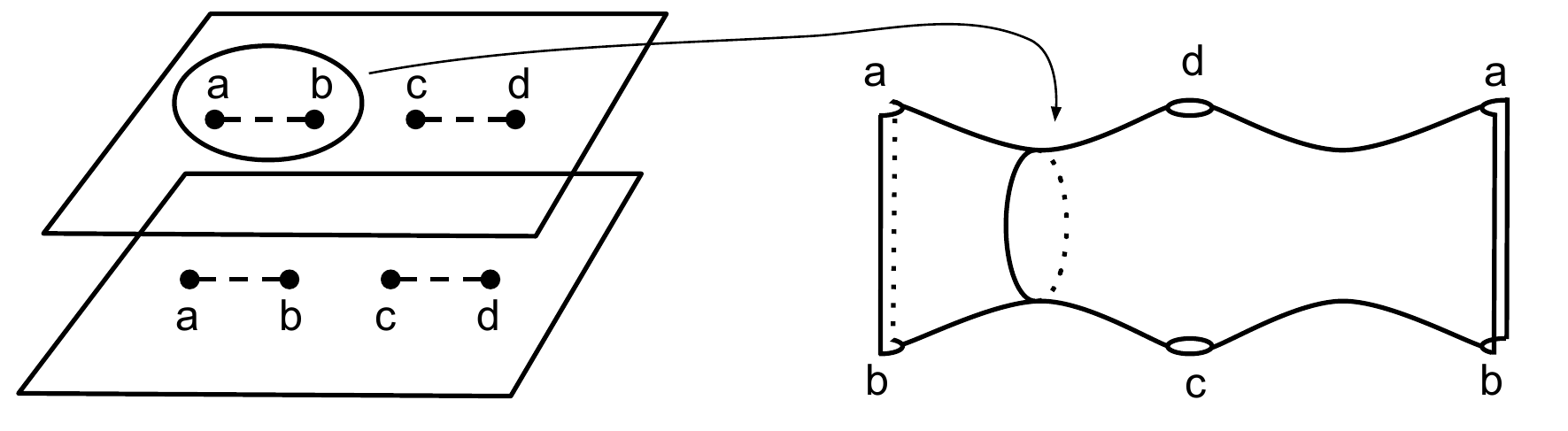}
\caption{\label{fig:defect} {\bf Defect worldvolume $\Sigma^{(1)}$}. When the torus is filled in, then this loop becomes contractible in the bulk. If the torus is considered as the boundary of its exterior, then this loop is not contractible in the bulk. This loop computes $\tr \left( \rho^{W_\Sigma} \rho \right)$.}
\end{figure}
In the previous section, in figure \ref{fig:final} and equations (\ref{eq:s1}) and (\ref{eq:sd}), we concluded that the exterior filling of the boundary geometry corresponded to the case in which the Ryu-Takayanagi surface was connected. We saw in equations (\ref{eq:orderone}) and (\ref{eq:infinite}) that the appropriate continuation of the function that was zero for $n>1$ and unity at $n=1$ has infinite derivative at $n=1$. If we furthermore assume that nonzero defect operator expectation values at integer $n$ analytically continue to have a finite derivative at $n = 1$ (as would seem to be generic -- close to $n=1$ the D-brane embedding should be entirely within one sheet of the bulk and therefore insensitive to the disappearance of the conical defect as $n\to 1$, see the appendix), then we can conclude that to leading order at large $N$:
\bea
S(\rho^{W_\Sigma} \parallel \rho) \; = \; \infty \qquad & \Leftrightarrow & \qquad \text{RT surface {\bf connected}.} \nonumber \\
S(\rho^{W_\Sigma} \parallel \rho) \; < \infty \qquad & \Leftrightarrow & \qquad \text{RT surface {\bf disconnected}.} \label{eq:final}
\eea
This is our first result. Recall that the relative entropy was defined in section \ref{sec:relative} above. There we also discussed the change in the entanglement entropy. The change in the entanglement entropy due to insertions of defect operators is finite, see e.g. \cite{Chang:2013mca, Jensen:2013lxa, Lewkowycz:2013laa, Karch:2014ufa}. The entropy change is therefore subleading in (\ref{eq:orderone}) and we can focus on the relative entropy. Also from (\ref{eq:orderone}) we see that the change in expectation value of the modular Hamiltonian is infinite when the relative entropy is infinite. This can perhaps be thought of as an entanglement analogue of the infinite energy cost to add a single external quark into a confined phase. The entanglement is `confining' relative to the defect operator insertion.

Applying the above logic to equations (\ref{eq:a}) and (\ref{eq:b}) we can also conclude that
\bea\label{eq:main2}
\text{$\rho^{W_\Sigma}$ and $\rho$ perfectly distinguishable} \quad & \Leftrightarrow & \quad \text{RT surface {\bf connected}.}  \nonumber \\
\text{$\hat \rho^{W_C}$ and $\hat \rho$ perfectly distinguishable} \quad & \Leftrightarrow  & \quad \text{RT surface {\bf discconnected}.} 
\eea
These are the main results of this paper.
This perfect distinguishability of a defect operator or Wilson loop insertion is a distinct -- gauge theoretic --  observable than the mutual information between the two disconnected regions. A nonzero mutual information and perfect distinguishability of the large $N$ density matrices have been tied together by gauge theoretic physics on topologically nontrivial backgrounds (arising in the computation of R\'enyi entropies).

Perfectly distinguishability implies infinite R\'enyi entropy. Furthermore, adapting the analytic continuation of the appendix to the density matrix $\hat \rho$ for the exterior rather than the interior of the regions $A$ (see section \ref{sec:wilson} above) we obtain the ``converse'' of (\ref{eq:final})
\bea
S(\hat \rho^{W_C} \parallel \hat  \rho) \; = \; \infty \qquad & \Leftrightarrow & \qquad \text{RT surface {\bf disconnected}.} \nonumber \\
S(\hat \rho^{W_C} \parallel \hat \rho) \; < \infty \qquad & \Leftrightarrow & \qquad \text{RT surface {\bf connected}.} \label{eq:final2}
\eea

\section{Final comments}

A connected Ryu-Takayanagi surface geometrizes the large $N$ mutual information between boundary regions. Our results link this fact to a statement about the Hilbert space of the theory. Namely, a geometrized (or not) mutual information is shown to be equivalent to two particular density matrices being perfectly distinguishable, with orthogonal support. This fact would seem to be of relevance for future microscopic formulations of the emergence of spacetime from entanglement. The r\^ole of R\'enyi entropies in our argumentation, placing the theory on spacetimes with nontrivial topology, is reminiscent of the characterization of topological order.

Codimension one defect operators have played a key r\^ole in our discussion. Similar codimension one defects have recently been proposed to quantify the complexity of holographic quantum states \cite{Stanford:2014jda}. The main point for us is that these operators can surround, in Euclidean spacetime, a component of a spatial region that is to be entangled. The discussion in \cite{Stanford:2014jda} involves space filling defects at a fixed instant in time in black hole geometries. As we have described, the $n=2$ R\'enyi entropy for two disconnected regions is the partition function of the theory on a space with topology $S^1 \times S^d$. The defect operators we have considered (surrounding one of the regions) lift to defects wrapping the $S^d$ in this space. These seem to be closely related to the defect insertions studied in \cite{Stanford:2014jda}.

Most of our topological derivations have effectively been via `proof by pictures'. It would clearly be desirable to have a more mathematically sophisticated treatment of these topics, especially for the higher dimensional case. This might allow a general result for general topology of the entangled spatial region (i.e. beyond the case of two disconnected disks that we have considered).

\section*{Acknowledgements}

It is a pleasure to thank Xi Dong, Matt Hastings, Jorge Santos and Douglas Stanford for helpful discussions on related topics. This work is partially supported by a DOE Early Career Award, by a Sloan fellowship and by a Gerhard Casper Stanford Graduate Fellowship.

\appendix

\section{Analytic continuation of defect operators}
\label{sec:continue}

In this appendix we give a holographic definition of the expectation value $\langle W_\Sigma^{(1)} \rangle_n/Z_n$ in (\ref{eq:us}) that is suitable for analytic continuation to non-integer $n$. The expectation value of such defect operators can be analytically continued following the ideas in \cite{Lewkowycz:2013nqa}. For integer $n$, the defect operator expectation value is given by the on-shell action of a classical D$(d+1)$-brane configuration $\Upsilon$ in the background spacetime dual to the $n$-fold cover boundary geometry. Thus
\be
\frac{\langle W_\Sigma^{(1)} \rangle_n}{Z_n} = e^{-S_{d+1}[\pa \Upsilon = \Sigma^{(1)}]} \,.
\ee
The bulk surface $\Upsilon$ corresponding to the D-brane worldvolume should not be confused with the Ryu-Takayanagi soap bubble. The relation $\pa \Upsilon = \Sigma^{(1)}$ means that the surface ends on the boundary theory defect operator worldvolume $\Sigma^{(1)}$.

\begin{figure}[h]
\centering
\includegraphics[height = 0.16\textheight]{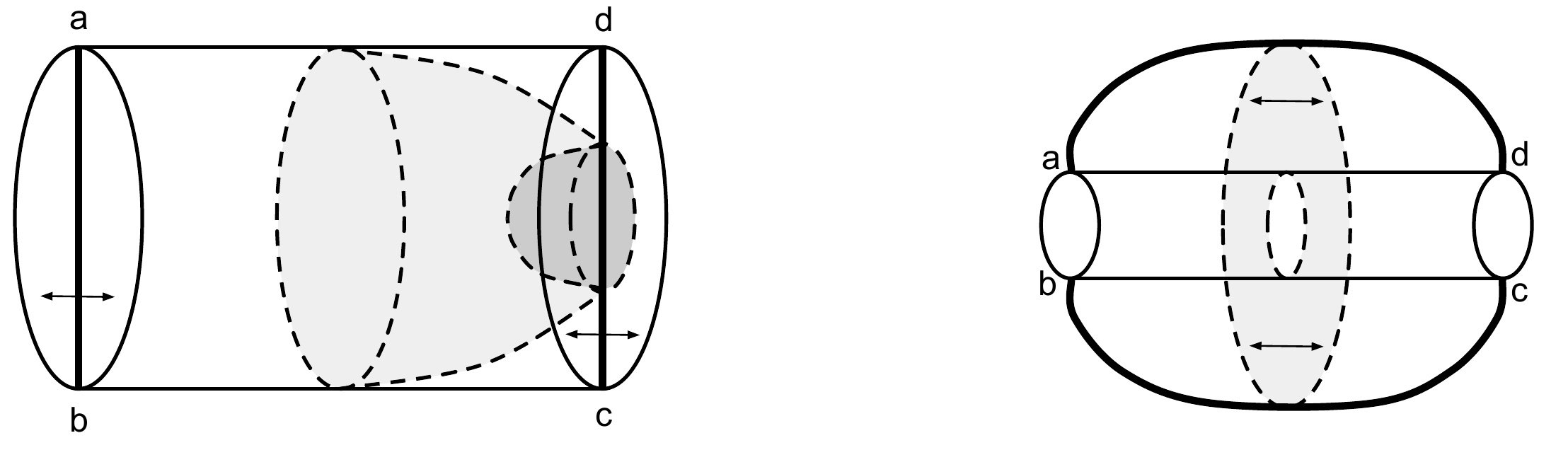}
\caption{\label{fig:m2} {\bf D-brane worldvolumes in $\widetilde M_2$}. The thick black lines are the fixed points of the $\Z_2$ action. Arrows denote the identification of would-be boundaries of $\widetilde M_2$ by the $\Z_2$ action (the only true interior boundaries are the lines of fixed points).
Left: $\widetilde M_2$ for the interior filling (cf Fig. \ref{fig:z2fill}). Right: $\widetilde M_2$ for the exterior filling (cf Fig. \ref{fig:z2fillout}). The light and dark grey components in the left plot together make out the full worldvolume. The darker region is obtained by projecting down the part of the original worldvolume on the second sheet of $M_2$. The candidate worldvolume in the right plot is different in that each intersection with the lines of fixed points involves only a single component (rather than two components in the left plot). This will be excluded by our boundary conditions at the conical defect.}
\end{figure}

Consider the case in which a D-brane worldvolume $\Upsilon$ satisfying the asymptotic boundary conditions exists in $M_n$. Then, because of the $\Z_n$ symmetry, we can simply project the worldvolume down to a worldvolume $\widetilde \Upsilon$ in $\widetilde M_n = M_n/\Z_n$. It is important that we include the entire surface in $M_n$ projected into $\widetilde M_n$. This is illustrated in figure \ref{fig:m2} below for the case of a 1+1 dimensional boundary.
The higher dimensional cases are analogous. Locally the background spacetime, D-brane worldvolume, and hence action are unchanged by this projection. The projected worldvolume can intersect the conical defect surface in $\widetilde M_n$ if the original worldvolume in $M_n$ crosses any of the bounding surfaces $S$ discussed in the previous section. This does not complicate the evaluation of the on shell action. In particular, we can recover the answer for the action by directly finding the worldvolume $\widetilde \Upsilon$ in $\widetilde M_n$ with the following boundary conditions: (i) $\widetilde \Upsilon$ must end on (one copy of) $\Sigma$ at the asymptotic boundary and (ii) at any intersection with the conical defect surface, there must be {\it two} components of the worldvolume meeting at the intersection. See the left hand part of figure \ref{fig:m2}, where the two components are shown in different shades of grey. Requiring two components to meet captures the fact that in the cover $M_n$ (for integer $n$), $\Upsilon$ continues on the other side of the bounding surface $S$. For such a surface $S_{d+1}[\pa \Upsilon = \Sigma^{(1)}]=S_{d+1}[\pa \widetilde \Upsilon = \Sigma]$. This reformulation is well defined for any $n$, including non-integer $n$, and therefore provides the required analytic continuation.

We must check that the analytic continuation of the previous paragraph has the property that when a D-brane doesn't exist in the full space $M_n$, then it also doesn't exist in the quotient space $\widetilde M_n = M_n/\Z_n$. These are the cases where the expectation value of the defect operator vanishes. For integer $n>1$, if a solution in existed in $\widetilde M_n$ -- in particular satisfying the boundary condition that two components of the worldvolume must meet at any intersection with the surface of conical defects -- then we could uplift it to a solution in $M_n$, obtaining a contradiction. Therefore a solution does not exist in $\widetilde M_n$. However, it is an assumption throughout our analysis that upon analytic continuation, the topology of $\widetilde M_n$, and in particular the topology of the conical defect surface in $\widetilde M_n$, does not change as a function of $n$ for $n>1$. Therefore the expectation value remains zero for all $n>1$, because a contractible bulk cycle on which we could wrap the D-brane does not exist. This is in agreement with the general expectation from (\ref{eq:infinite}). In the right hand side of figure \ref{fig:m2} we illustrate how D-brane worldvolumes that might seem to exist in this case do not satisfy the condition of having two components meet at intersections with the conical defects. At $n=1$, there is no conical defect surface. There is no topological constraint in this case. This is again in agreement with the general discussion around equation (\ref{eq:infinite}) above.

The above method can also be used to analytically continue the Wilson loop expectation value $\tr  \left( \hat \rho^{W_C} \hat \rho^{n-1} \right)$ discussed in section \ref{sec:wilson}.

\end{document}